\documentclass[aps,pra,twocolumn,showpacs,superscriptaddress]{revtex4-1}
\usepackage[latin1]{inputenc}
\usepackage{graphicx}
\usepackage{epstopdf}
\usepackage{epsfig}
\usepackage{xfrac}
\usepackage{xcolor}
\usepackage{indentfirst}
\usepackage{amsmath, amsthm, amssymb} 
\usepackage{bm}
\usepackage{float}
\usepackage[colorlinks=true, a4paper=true, pdfstartview=FitV, linkcolor=blue, citecolor=blue, urlcolor=blue]{hyperref}
\usepackage{braket}
\begin{document}  
\title{Roton-Induced Trapping in Strongly Correlated Rydberg Gases} 
\author{Jo\~{a}o D. Rodrigues}
\email[Corresponding author:]{joaodmrodrigues@tecnico.ulisboa.pt}
\affiliation{Instituto de Plasmas e Fus\~{a}o Nuclear, Instituto Superior T\'{e}cnico, Universidade de Lisboa, 1049-001 Lisbon, Portugal}
\author{Lu\'{i}s F. Gon\c calves}
\affiliation{Rydberg Technologies LLC, Ann Arbor, Michigan 48103, USA}
\author{Hugo Ter\c cas}
\affiliation{Instituto de Plasmas e Fus\~{a}o Nuclear, Instituto Superior T\'{e}cnico, Universidade de Lisboa, 1049-001 Lisbon, Portugal}
\author{Lu\'{i}s G. Marcassa}
\affiliation{Instituto de F\'{i}sica de S\~{a}o Carlos, Universidade de S\~{a}o Paulo, Caixa Postal 369, 13560-970, S\~{a}o Carlos, S\~{a}o Paulo, Brasil}
\author{Jos\'{e} T. Mendon\c ca}
\affiliation{Instituto de Plasmas e Fus\~{a}o Nuclear, Instituto Superior T\'{e}cnico, Universidade de Lisboa, 1049-001 Lisbon, Portugal}
\begin{abstract}
\begin{footnotesize}
Atoms excited into high-lying Rydberg states and under strong dipole-dipole interactions exhibit phenomena associated with highly correlated and complex systems. We perform first principles numerical simulations on the dynamics of such systems. The emergence of a roton minimum in the excitation spectrum, as expected in strongly correlated gases and accurately described by Feynman's theory of liquid helium, is shown to significantly inhibit particle transport, with a strong suppression of the diffusion coefficient, due to the emerging spatial order. We also demonstrate how the ability to temporally tune the interaction strength among Rydberg atoms can be used in order to overcome the effects of disorder induced heating, allowing the study of unprecedented highly coupled regimes.
\end{footnotesize}
\end{abstract}
\maketitle
%
\section{Introduction}
\par
The concept of the roton minimum was first introduced by Landau in an attempt to explain superfluidity in liquid helium \cite{Landau1941, Landau1947}. Despite its rather heuristic derivation, and the fact that its microscopical origin has yet to be fully understood, Landau's phenomenological model accounted for many of the experimental features \cite{Palevsky1958, Henshaw1958, Dietrich1972, Griffin2009}. Later, Feynman provided a deeper understanding on the physical origin of the roton minimum by relating it with the structural properties of the superfluid state \cite{Feynman1954, Feynman1956}, particularly the presence of strong spatial correlations and the high-interactive character among helium atoms. Roton-like excitations have also been observed in different systems, such as quantum Fermi liquids \cite{Godfrin2012} and dipolar Bose-Einstein condensates \cite{Santos2003, Santos2009, Chomaz2018}, while roton instabilities have been related with the formation quantum supersolids in spin-orbit coupled \cite{Li2017} and Rydberg-excited Bose-Einstein condensates \cite{Henkel2010}, as well as in cold atoms above degeneracy due to multiple scattering of light \cite{Tercas2012}.
\par
Recently, Rydberg gases \cite{gallagher, Marcassa2014} have also been the subject of great scientific interest, mainly due to their high interactive character. One such manifestation of strong interactions is the so-called blockade effect, either due to van der Waals \cite{Tong2004} or dipole-dipole forces \cite{Urban2009}. Strong interactions among Rydberg atoms play two distinct, albeit related, roles. On the one hand, the dynamics and transport of Rydberg excitations \cite{Schausz2012, Labuhn2016} develops at fast timescales, where the motional degrees of freedom are essentially frozen. On the other hand, at slower timescales, the mechanical (external degrees of freedom) evolution of Rydberg gases \cite{McQuillen2013, Thaicharoen2015, Thaicharoen2016, Faoro2016} present a great approach to the physics of strongly interacting and highly correlated particle systems, a subject of extensive theoretical, computational and experimental research \cite{Ichimaru1982, Bollinger1984, Chu1994, Dubin1999, Knapek2007, Ashwin2010, Ashwin2011, Bannasch2012}. 
\par
In this paper, we perform molecular dynamics (MD) simulations of two-dimensional Rydberg gases. A roton minimum in the excitation spectrum is shown to emerge in the strongly coupled regime and is accurately described by Feynman's theory of liquid helium, relating it to the development of spatial ordering. The origin of the roton minimum and its relation with other systems is discussed in detail. By tracking the trajectory of each atom and studying its mean square displacement we show that the emerging spatial correlations are responsible for a strong inhibition of (diffusive) transport, as atoms become trapped in the roton minima. We also investigate the relaxation dynamics and describe the mitigation of the long-standing problem of disorder induced heating. This is accomplished by adiabatically increasing the interaction strength during thermalization, culminating in the achievement of stronger correlated samples. The experimental realization of the phenomenology described here is also discussed. 
%
%
%
%
\section{Theoretical Model}
\par
Permanently polarized Rydberg atoms interact, in a first approximation, via the pair-wise dipole-dipole potential \cite{Jackson, Reinhard2007, Comparat2010, Thaicharoen2016}
\begin{equation}\label{eq:dipole_potential}
V  (r, \theta ) = \frac{C_3}{r^3} \left(1 - 3\cos^2  (\theta ) \right),
\end{equation}
with $\theta$ the angle between the atomic dipoles $\mathbf{P}$ (aligned with external electric field $\mathbf{E}_0$) and the interatomic separation vector $ \mathbf{r}$. The strength of the interaction is quantified by $C_3 =  P^2 / 4 \pi \epsilon_0$, where $P$ denotes the induced permanent atomic dipole moment and $\epsilon_0$ the vacuum electric permitivity. While in one dimension the interaction is always isotropic, with attractive or repulsive character depending on the polarization angle $\theta$ \cite{Goncalves2016}, in a two-dimensional gas isotropic repulsion is obtained with polarization perpendicular to the atomic sample. In this case, $\theta = \pi / 2$ and the in-plane potential is reduced to $V(r ) = C_3 / r^3$, which is the case we consider throughout this paper. In a two-dimensional sample we define the average inter-particle distance (Wigner-Seitz radius) as $a=(\pi n_0)^{-1/2}$, with $n_0 = 1/\pi$ the homogeneous density, in the units of $r/a$. The degree of correlations in the gas can be properly parametrized by the coupling parameter, defined as the mean ratio of potential to kinetic energy, $\Gamma = \langle V(r) \rangle / k_B T$, with $k_B$ the Boltzmann constant and $T$ the temperature of the gas. In terms of the mean inter-particle distance $\Gamma = C_3 /k_B T a^3$, with weakly and strongly correlated regimes corresponding to $\Gamma \ll 1$ and $\Gamma \gg 1$, respectively.
\par
In order to describe the emergence of spatial correlations (expected for $\Gamma > 1$), the radial distribution function $g(r)$ is defined such that the average number of atoms between $r$ and $r+dr$ from a reference atom is $2 \pi r n_0 g(r) dr$. In the absence of correlations $g(r) = 1$, or $h(r) = 0$, with $h(r) = g(r) -1$ the pair correlation function. In reciprocal space we may also define the (static) structure factor, related with $g(r)$ via $S(k) -1 = n_0 \int d \mathbf{r} e^{-i \mathbf{k} \cdot \mathbf{r}} \left[ g(r) -1 \right]$. Both these structural function can be obtained, even in the strongly coupled regime, via the integral equation technique and the hyper-netted chain (HNC) closure equation \cite{Rodrigues2018} - check Appendix \ref{sec:appendixA} for further details.
\par
The correlational origin of the roton minimum can be traced back to the early work of Feynman regarding the structural properties of liquid helium. At finite temperatures, Feynman's theory relates the static structure factor with the dispersion relation of the elementary excitation $\omega(k)$ via \cite{Feynman1954}
\begin{equation}\label{eq:feynman_1}
S(k) = \frac{\hbar k^2}{2m \omega(k)} \text{coth} \left( \frac{ \hbar \omega(k)}{2 k_B T} \right) \simeq \frac{u_s^2 k^2}{\omega^2(k)},
\end{equation}
with $u_s = \sqrt{k_B T / m}$ the thermal speed of sound. The last expression is obtained by taking the classical limit $k_B T \gg \hbar \omega$, and corresponds to the result obtained via the classical fluctuation-dissipation theorem \cite{Wang1997}. This is a further evidence of the close connection between the roton minimum and classical correlations induced by strong interactions. It also goes in line with previous arguments \cite{Kalman2010}, where quantum and classical simulations of strongly correlated systems are shown to yield equivalent results regarding the excitation spectrum, discarding the influence of quantum fluctuations, suggesting similar physics with that of liquid Helium \cite{Nozieres2004}. The connection between the spatial correlations at high $\Gamma$ and the dynamics of the gas, as determined by Feynman's model, is further illustrated in the Appendix \ref{sec:appendixB}. As an important remark, notice that the (correlational) roton minimum is of a distinct physical origin as that in the case of dipolar Bose-Einstein condensates \cite{Santos2003, Santos2009}. In the later, the roton minimum in quasi-2D samples depends on the slight three-dimensional character of the dipole-dipole potential, in the weakly interacting regime, while here the roton minimum emerges in the strongly correlated limit even in pure 2D samples.
%
%
%
\section{Molecular dynamics and emergence of the roton minimum}
\par
In the sequence, we perform Molecular Dynamics (MD) simulations of the two dimensional Rydberg gas, with $N=5000$ atoms distributed in a square box of size $L=\sqrt{N/n_0}=\sqrt{\pi N} \simeq 125a$, much larger than the typical range of spatial correlations, at least in the range of parameters investigated here. This is a critical feature and ensures the absence of spurious correlations due to finite size effects. Periodic boundary conditions are used to mimic the evolution of an homogeneous and isotropic system. In the early simulation times, we apply a thermalization algorithm where the particles velocities are rescaled accordingly as to maintain high values of the coupling parameter. To this end, before each iteration of a regular velocity Verlet algorithm we calculate the mean velocity over the entire collection of particles, typically initiated such that $\langle v \rangle = 1$. After each integration step the updated velocities are normalized in order to maintain the mean velocity of the entire system. This can be understood as a relaxation at constant temperature - canonical evolution. Once thermalization has been settled, here identified by the convergence of the velocity distribution, the trajectories are integrated in a full Hamiltonian manner - microcanonical evolution. Distances are measured in units of the mean interparticle distance $a$, time in units of $a/u_s$ and velocities normalized to the thermal speed of sound $u_s$. 
\begin{figure}
\centering
\includegraphics[scale=0.620]{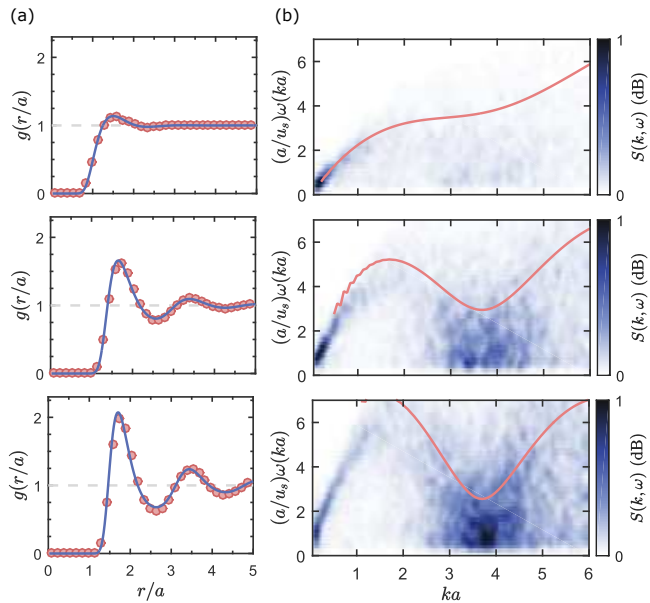}
\caption{Results of Molecular Dynamics simulation for a two-dimensional Rydberg gas. From top to bottom we increase the coupling parameter ($\Gamma = 2, 15, 35$) and observe the emergence of strong spatial correlations and the resulting roton minimum in the spectrum of density fluctuations. (a) Radial distribution function obtained from the MD simulation (red dots) and comparison with the HNC prediction (blue line). (b) Dynamic structure factor obtained from the MD trajectories and comparison with Feynman's model of liquid helium based on the HNC correlations (red line) showing the emergence of a roton minimum near $ka \sim 4$.}
\label{fig:figure2}
\end{figure}
\par
The spectrum of density fluctuations (dynamic structure factor) is obtained from the atomic trajectories (after thermal equilibrium has been reached) by defining the intermediate scattering function as $\rho(k,t) = \sum_i \text{exp} \left[i \mathbf{k} \cdot \mathbf{r}_i (t) \right]$ \cite{Donko2008, simple_liquids}, where the summation is taken over all particles, with the dynamic structure factor following as \cite{Hansen1975}
\begin{equation}\label{eq:DSF}
S(k,\omega) = \frac{1}{2 \pi N} \lim_{{\Delta T \rightarrow \infty}} \frac{1}{\Delta T} \lvert \rho(k ,\omega) \rvert^2,
\end{equation}
with $\Delta T$ the total simulation time and $\rho(k, \omega)$ the Fourier transform of the intermediate  scattering function. From the symmetry of the dipole-dipole potential, we assume that $S(k,\omega)$ does not depend on the direction of the perturbation wavevector. The time is discretized in steps of $0.01$ with a total simulation span of 50, in units of $a/u_s$. Such long simulation times ensure high frequency resolution upon Fourier transformation. Also, permissible wavenumbers are given by multiples of the inverse box size, namely $k_\text{min} = 2 \pi / L$. Collective excitation are identified as peaks in the spectrum of density fluctuations, with their respective width related with the excitation lifetime - narrower peaks entail longer-lived modes. 
\par
The results of the MD simulations are presented in Fig. (\ref{fig:figure2}). The mode softening observed near $ka \sim 4$ entails the emergence of a roton minimum, accompanied with the development of short range order in the spatial arrangement of atoms. Also, the integral equation technique (and HNC closure) prediction accurately describes the observed spatial ordering quantified by the radial distribution function. Notice the increasing absence of particles - $g(r) \simeq 0$ - near $r=0$ for increasing values of the coupling parameter. This is due to the stronger dipole-dipole repulsion at higher values of $\Gamma$. The radial distribution function, together with Feynman's model of liquid helium in Eq. (\ref{eq:feynman_1}) correctly describes the MD results, with an excellent agreement observed without any free fitting parameter. The blurring of the $S(k, \omega)$ near $k_r$ indicates relative short lifetime the excitations near the roton position, at least in comparison with the longer-lived modes in the acoustic region near $k \rightarrow 0$. As an important remark, notice the increasing noise displayed by the theoretical dispersion relation (red line) in the low $k$ range. This is due to the finite size of the simulation box. For increasing value of $\Gamma$, the extent of spatial correlations increases. Upon Fourier transformation of the radial distribution function, in order to obtain the static structure factor and, consequently, the dispersion relation according to Feynman's theory, these finite size effects lead to the emergence of noisy features in the low $k$ range of $S(k)$. The correlational origin of the roton minimum has a simple physical reasoning. The mode softening near the roton momentum at $k_r$ indicates a slowing down of the phase velocity of density perturbations, $\omega(k_r) / k_r$, allowing for the accumulation of atoms near $2\pi / k_r$ and resulting in peaks in radial distribution function at this periodicity.
%
%
%
%
\begin{figure}
\centering
\includegraphics[trim={0 0cm 0 0cm}, clip, scale=0.75]{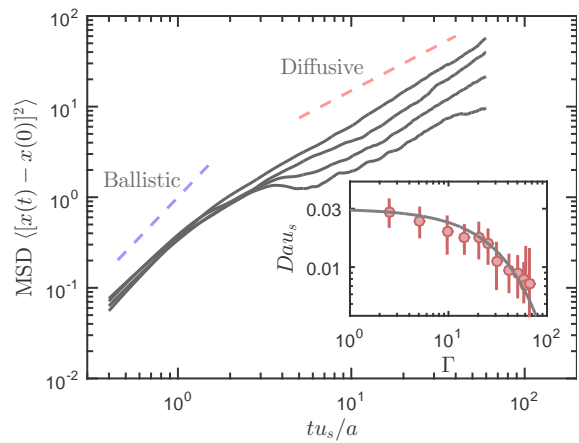}
\caption{Diffusion in a strongly interacting Rydberg gas. Mean-square displacement (in the $x$ direction) as a function of time, for different values of the coupling coefficient - from top to bottom $\Gamma = 2, 10, 20, 40$. The results for the $y$ direction are equivalent. The inset depicts the diffusion coefficient as a function of $\Gamma$, with strong transport inhibition in the highly interacting regime. The solid line theoretical curve is based the Einstein-Stokes relation together with the Arrhenius-like law $\eta \sim e^{0.025\Gamma}$.}
\label{fig:figure3}
\end{figure}
\section{Inhibition of diffusion}
\par
The (diffusion) transport properties follow from the evaluation of the mean-square displacement (MSD) of individual atoms, defined as (in the $x$ direction, for instance) $\langle \left[ x(t) - x(0) \right]^2 \rangle$. Pure diffusion (Brownian motion) is characterized by the Einstein relation $\text{lim}_{t \rightarrow \infty} \langle \left[ x(t) - x(0) \right]^2 \rangle = 2 D t$, with $D$ the diffusion coefficient. The simulation results are presented in Fig. (\ref{fig:figure3}). We observe ballistic-like motion in the early simulation times, coincident with the initial relaxation at constant temperature, as described before. Once thermalization settles, pure diffusive transport takes place. Here, the simulation trajectories are taken over the microcanonical ensemble, corresponding the most relevant experimental situation. The mean-square displacement allows the obtainment of the diffusion equation directly from the Einstein relation. Here, we observe a strong inhibition of diffusion (transport) in the strongly interacting regime. This is a rather distinct, albeit related, manifestation of the emerging spatial order and the corresponding roton minimum, as dipole-dipole interactions ensure the transient ``trapping'' of atoms at local minima in the potential landscape, as described by the radial distribution functions described before. As an import remark, notice the increasing error bars associated with the diffusion coefficient. Although an overall diffusive behavior is always observed, there is an increasing irregular behavior of the mean-square displacement at higher values of $\Gamma$. 
\par
In the presence of shear-viscosity, here induced by dipole-dipole interactions, diffusion can be described by the Einstein-Stokes relation $D = k_B T / q a \eta$, with $\eta$ the shear-viscosity coefficient and $q$ a numerical factor dependent on intrinsic properties of the medium \cite{Zwanzig1983}. Moreover, when transient trapping dominates the transport in strongly correlated systems, the shear-viscosity can be determined by an Arrhenius-type relation $\eta = A e^{B \Gamma}$ \cite{Daligault2006}, with $A$ and $B$ numerical factors. This type of relation can be justified by Eyring's theory of transport in liquids \cite{Eyring1936}. Altogether, this yields a dependence of the form $D \sim e^{-\Gamma}$, which fits well with the MD results - see Fig. (\ref{fig:figure3}).
%
%
%
%
\section{Experimental considerations and disorder induced heating}
\par
The experimental excitation of highly interacting Rydberg atoms precludes the obtainment of large densities and overall coupling strengths, due to the limitations imposed by the blockade mechanism. As such, an excitation into a low interacting n$S_{1/2}$ state, for instance, ensures small interatomic separations. In this case, the blockade effect only arises from the weaker van der Waals (repulsive) interaction, which becomes significant only for higher $n$ (principal quantum number). In a recent work \cite{Goncalves2016}, an one-dimensional sample of $^\text{85}$Rb atoms confined in an optical dipole trap with density $10^{12}$ cm$^{-1}$ and temperature of $80$ $\mu$K was used to measure the angular dependence of the dipole interaction by applying an external electric field. Here, the authors directly excited Rydberg atoms in $S$-like states between two avoided-crossings in the Stark manifold and measured an interaction parameter $C_3$ of approximately $99.7$ MHz.$\mu$m$^3$ and an interatomic distance of $\sim$ $4$ $\mu$m. Such high interaction parameter for $S$ states was recently confirmed by preliminary calculations \cite{Jim2018}. This simple atomic system carries a relatively high coupling parameter $\Gamma \sim 10$.
\par
An effective way to overcome the Rydberg atom density-limit in highly interacting states (imposed by the blockade effect), and achieve even higher coupling parameters consists in performing a Landau-Zener adiabatic passage, promoted by an electric field sweep through an avoided crossing in the Stark landscape. This transfers the atoms from $nS$ into high dipolar states, which become permanently polarized in the direction of the electric field, moving the atomic sample into an highly interacting regime \cite{saquet2010, wang2015, Thaicharoen2016, Rodrigues2018}. 
\begin{figure}
\centering
\includegraphics[trim={0 0cm 0 0cm}, clip, scale=0.75]{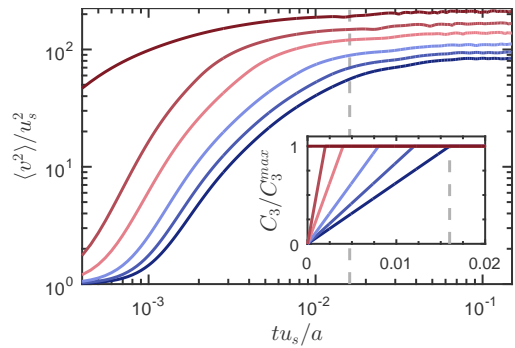}
\caption{Thermalization of a two-dimensional Rydberg gas with time-dependent interaction strength. Average square-velocity during relaxation of the initially disordered gas for different time-dependent interaction strength, as depicted in the inset plot. The dashed-line corresponds to the same time instant and the different colors to the same conditions on both plots.}
\label{fig:figure4}
\end{figure}
\par
Immediately after excitation, where no particular order is expected, there is a rapid reorganization of the atoms into an equilibrium distribution. Here, the randomly positioned dipoles are accelerated by the potential landscape, which is accompanied by a fast increase of the temperature and the decrease in the overall coupling coefficient. This is essentially the mechanism of disorder induced heating (DIH) that has been reported in ultra cold neutral plasmas (UCNPs) \cite{Killian2007}. The amount of heating is directly related with the level of disorder (lack of correlations) and, the excitation of initially ordered states should greatly reduce these effects. Proposals to prepare such initially ordered states include taking advantage of the Rydberg blockade mechanism \cite{Bannasch2013}, or via the preordering of atoms in an partially filled optical lattice \cite{Murphy2016}. While the latter could also be employed in the present context, the former more simply means that an initial excitation where the interatomic distance is limited by the blockade mechanism would also suppress the effects associated with disorder induced heating.
\par
Here, we further demonstrate that, the ability to temporally control the interaction strength (quantified by $C_3$) via the external electric field, can be exploited in order to partially overcome the effects of DIH and reach stronger coupled equilibrium samples. The key idea is to slowly increase the interaction strength during the relaxation process, in such a way that the atoms are allowed to adiabatically follow the changing potential landscape while adapting to the transient equilibrium configurations. The numerical results are portrayed in Fig. (\ref{fig:figure4}). We observe that slower ramping of the interaction strength results in lower temperature, and hence, higher coupled equilibrium gases, partially mitigating the heating effect due to the initial disorder. Moreover, for a sample of $^\text{85}$Rb at a temperature of 100 $\mu$k and $a\sim 2$ $\mu$m, as is the case of typical experiments \cite{Thaicharoen2016}, we obtain $a / u_s \sim 20$ $\mu$s. Hence, ramping times even slower than those considered in Fig. (\ref{fig:figure4}), should experimentally lead to thermalization at high values of $\Gamma$ and at timescales much shorter than both the typical lifetime of Rydberg states and also the time range available to recover the atomic trajectories \cite{Thaicharoen2016}, pointing towards the feasibility of the experimental observation of the roton minimum, as well as other features associated with strongly coupled system, including Wigner crystallization \cite{Tan1995, Pohl2004}
%
%
\section{Conclusion}
\par
We have performed MD simulations of two dimensional gases of highly interacting Rydberg atoms and demonstrated the emergence of a roton minimum in the excitation spectrum accompanied by the development of strong spatial correlations. A model based on Feynman's theory of superfluidity was shown to accurately describe the numerical results, evidencing the correlational origin of the roton minimum, as first anticipated for liquid helium, and here demonstrated for a system under classical (dipole) interactions. The transient trapping of atoms in the ``roton potential" was shown to strongly inhibit diffusive transport. We also demonstrated that an adiabatic ramping of the interaction strength during the initial relaxation of the gas, easily achievable in typical experiments involving Rydberg atoms, can be used to highly mitigate the effects of disorder induced heating \cite{Thaicharoen2016, Killian2007, Bannasch2013}, offering a route towards experimental investigation of otherwise inaccessible regimes of strongly coupled gases.
%
%
%
%
\section*{Acknowledgements}
\par
This work was partially supported by S\~{a}o Paulo Research Foundation (FAPESP) Grants No. 2011/22309-8 and No. 2013/02816-8, the U.S. Air Force Research Grant FA9550-16-1-0343  and CNPq. LFG acknowledges the support of Rydberg Technologies LLC. Also, HT acknowledges FCT - Funda\c{c}\~{a}o para a Ci\^{e}ncia e Tecnologia (Portugal) - through Grant No. IF/00433/2015.
%
%
%
%
\appendix
\section{Integral Equation Technique \label{sec:appendixA}}
\par
In the study of strongly correlated systems, one must go beyond hydrodynamical or Vlasov descriptions in terms of single-particle densities. The full $N$-particle density, $ \rho^{(N)} ( \mathbf{r}^N )$, is usually truncated to the lower order single, and the two particle functions, $\rho^{(1)} \left( \mathbf{r}\right) =  \langle \sum_i \delta(\mathbf{r} - \mathbf{r}_i ) \rangle$ and $\rho^{(2)} ( \mathbf{r}, \mathbf{r'} ) = \langle \sum_i \sum_{j \neq i} \delta(\mathbf{r} - \mathbf{r}_i )  \delta(\mathbf{r'} - \mathbf{r}_j ) \rangle$, respectively, with summations over the entire set of $N$ particles and averaging over the canonical ensemble, usually. With the single particle density $\rho^{(1)}(\mathbf{r}) = n_0$, correlations are usually described by the (normalized) two-particle distribution $g^{(2)}(\mathbf{r}_1, \mathbf{r}_2) = \rho^{(2)}(\mathbf{r}_1, \mathbf{r}_2) / n_0^2$. For isotropic interactions $g^{(2)}(\mathbf{r}_1, \mathbf{r}_2) = g^{(2)}( \lvert \mathbf{r}_1 -  \mathbf{r}_2 \rvert) \equiv g(r)$, the latter known as the radial distribution function, defined such that the average number of particles between $r$ and $r+dr$ away from an atom at $r=0$ is $2 \pi r n_0 g(r)dr$ (in two-dimensional samples).
\par
Apart from the radial distribution function, $g(r)$, the structural properties of a gas can be inferred from a dual description in reciprocal space by defining the (static) structure factor as
\begin{equation}\label{eq:structure_factor0}
S(\mathbf{k}) = \frac{1}{N} \langle \rho^{(1)} (\mathbf{k}) \rho^{(1)} (\mathbf{-k}) \rangle = 1 + \frac{1}{N} \langle \sum_{i \neq j} e^{-i \mathbf{k} \cdot \left( \mathbf{r}_i - \mathbf{r}_j \right)} \rangle,
\end{equation}
with $S(\mathbf{k}) = 1$ indicating the absence of correlations. The radial distribution function and the static structure factor are related by
\begin{equation}\label{eq:structure_factor3}
S(k) -1 = n_0 \int d \mathbf{r} e^{-i \mathbf{k} \cdot \mathbf{r}} \left[ g(r) -1 \right],
\end{equation} 
or, equivalently, $S(k) = 1+n_0 h(k)$. 
\par
Correlations between a pair of Rydberg atoms arise both from the direct and indirect interactions. In this context, we may introduce the direct correlation function $c(r)$, related with the total correlation $h(r)$ by the Ornstein-Zernike relation \cite{simple_liquids}
\begin{equation}\label{eq:OZ1}
h(r) = c(r) + n_0 \int d\mathbf{r'} c( \lvert \mathbf{r} - \mathbf{r'} \rvert) h(r').
\end{equation}
with $h(r)$ including the effect of direct and indirect correlations, as those originating from interactions between any pair of intermediate particles. As such, while the range of $c(r)$ is usually comparable with that of the pair potential, the total correlation function is of higher range, due to the effects of indirect correlations.
\par
In order to close the OZ relation, we begin by noting that the distribution of particles in the presence of interactions may be given according to the Boltzmann relation, namely
\begin{equation}\label{eq:barometric}
n(r) = n_0 \text{exp} \left[ -\phi(r)/k_B T \right],
\end{equation}
and known as the barometric law. Here, $\phi(r)$ is the total potential and can be constructed in a hierarchy similar to the OZ relation, where the total potential, $\phi(r)$, is the sum of the direct pair-wise term, $V(r)$, and the indirect contribution from any number of intermediate atoms, namely
\begin{equation}\label{eq:hnc_explanation}
\begin{split}
- \frac{\phi(r)}{k_B T} & =  -\frac{V(r)}{k_B T} + n_0 \int d\mathbf{r'} c( \lvert \mathbf{r} - \mathbf{r'} \rvert) h(r') \\
& = - \frac{V(r)}{k_B T} + h(r) - c(r),
\end{split}
\end{equation}
where the last equality follows from the OZ relation. From the barometric law in Eq. (\ref{eq:barometric}) follows
\begin{equation}\label{eq:hnc}
g(r) = \text{exp} \left[ - \frac{V(r)}{k_B T} + h(r) - c(r) \right],
\end{equation}
also known as the hyper-netted chain (HNC) closure relation \cite{simple_liquids}. Together with the OZ relation in Eq. (\ref{eq:OZ1}) forms a closed set of equations for the total and direct correlations functions, $h(r)$ and $c(r)$, respectively. The details on the numerical algorithm for the solution of the coupled Eqs. (\ref{eq:OZ1}) and (\ref{eq:hnc}) can be found in Ref. \cite{Rodrigues2018}.
%
%
%
%
\appendix
\section{Correlational Origin of the Roton Minimum \label{sec:appendixB}}
\par
\begin{figure}
\centering
\includegraphics[trim={0 0cm 0 0cm}, clip, scale=0.70]{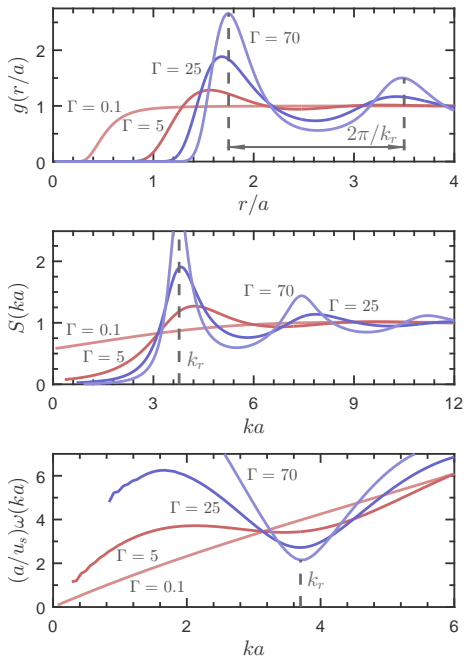}
\caption{Two-dimensional Rydberg gas. Top-panel: Radial distribution function and emergence of short-range correlations in the strongly coupled regime ($\Gamma \gg 1$), with $k_r$ the wavenumber associated with the regularity in the spatial arrangement. Middle-panel: Static structure factor, with the developing peak at approximately $k_r = 2 \pi / \lambda_r \simeq 4a^{-1}$ and related with the short-range oscillatory behavior of the radial distribution function. Bottom-panel: Excitation spectrum of the atom density fluctuations and the emergence of a roton minimum near $k_r$.}
\label{fig:figure1}
\end{figure}

The correlational origin of the roton minimum can be understood from Feynman's model presented in the main text or, equivalent in the classical limit, the relation between the static structure factor and the dispersion relation as dictated by the fluctuation-dissipation theorem \cite{Wang1997} - Fig. (\ref{fig:figure1}). Here, the radial distribution function - top panel - and the (static) structure factor - middle panel - are numerically obtained from the HNC relation \cite{Rodrigues2018}. We observe the emergence of a roton minimum at $k_r$ - bottom panel - associated with the developing short-range order in the spatial arrangement of the Rydberg atoms, at a periodicity of approximately $2\pi / k_r$. 
\bibliographystyle{apsrev4-1}
\bibliography{references}
\bibliographystyle{apsrev4-1}
\end{document}